\documentclass[sigconf]{acmart}
\AtBeginDocument{%
  \providecommand\BibTeX{{%
    \normalfont B\kern-0.5em{\scshape i\kern-0.25em b}\kern-0.8em\TeX}}}

\setcopyright{acmlicensed}
\copyrightyear{2018}
\acmYear{2018}
\acmDOI{XXXXXXX.XXXXXXX}

\acmConference[CHI '24]{Proceedings of the CHI Conference on Human Factors
in Computing Systems}{May 11--16, 2024}{Honolulu, HI, USA}
\acmBooktitle{Proceedings of the CHI Conference on Human Factors in
Computing Systems (CHI '24), May 11--16, 2024, Honolulu, HI, USA}
\acmISBN{978-1-4503-XXXX-X/18/06}

\definecolor{brown}{rgb}{0.59, 0.29, 0.0}
\definecolor{darkgray}{rgb}{0.59, 0.59, 0.59}
\definecolor{tablegray}{gray}{.9}

\usepackage[utf8]{inputenc}
\usepackage{diagbox}
\usepackage{colortbl}

\usepackage{tabularx}

\usepackage{color}
\usepackage{enumitem}
\usepackage{mathtools}
\usepackage{commath}

\usepackage{amssymb}
\usepackage{pifont}

\newcommand{\customtilde}{{\raise.17ex\hbox{$\scriptstyle\sim$}}}

\usepackage{xparse}

\setcopyright{none}

\renewcommand\footnotetextcopyrightpermission[1]{}

\begin{document}

\title{Deceptive Patterns of Intelligent and Interactive Writing Assistants}


\author{Karim Benharrak}
\orcid{0009-0002-3279-5664}
\email{karim@benharrak.com}
\affiliation{%
  \institution{The University of Texas at Austin}
  \city{Austin}
  \state{TX}
  \country{USA}
}

\author{Tim Zindulka}
\orcid{0009-0009-1972-351X}
\email{tim.zindulka@uni-bayreuth.de}
\affiliation{%
  \institution{University of Bayreuth}
  \streetaddress{Universitätsstr. 30}
  \city{Bayreuth}
  \state{Bavaria}
  \country{Germany}
  \postcode{95447}
}

\author{Daniel Buschek}
\orcid{0000-0002-0013-715X}
\email{daniel.buschek@uni-bayreuth.de}
\affiliation{%
  \institution{University of Bayreuth}
  \streetaddress{Universitätsstr. 30}
  \city{Bayreuth}
  \state{Bavaria}
  \country{Germany}
  \postcode{95447}
}

\renewcommand{\shortauthors}{Benharrak et al.}

\begin{abstract}
Large Language Models have become an integral part of new intelligent and interactive writing assistants.
Many are offered commercially with a chatbot-like UI, such as ChatGPT, and provide little information about their inner workings. 
This makes this new type of widespread system a potential target for deceptive design patterns. For example, such assistants might exploit \textit{hidden costs} by providing guidance up until a certain point before asking for a fee to see the rest. As another example, they might \textit{sneak} unwanted content/edits into longer generated or revised text pieces (e.g. to influence the expressed opinion).
With these and other examples, we conceptually transfer several deceptive patterns from the literature to the new context of AI writing assistants. Our goal is to raise awareness and encourage future research into how the UI and interaction design of such systems can impact people and their writing.
\end{abstract}





\maketitle

\section{Introduction}
In this short paper, we transfer known deceptive UI/UX patterns~\cite{Gray2018} to the context of interactive and intelligent writing assistant systems \cite{Lee2024dsiiwa}. 

Our goal is to raise awareness of the potential use of such patterns in recently popular applications of Large Language Models (LLMs) in interactive systems, such as ChatGPT\footnote{\url{https://chat.openai.com/}} and systems that offer (AI) assistance for text-related tasks. We do \textit{not} claim that these patterns are used in specific products at the time of writing. That said, we have anecdotally found examples that are very close to what such deceptive patterns could look like in this context.

In general, deceptive patterns in UI/UX design are design choices that deliberately deceive users, often to increase profit: Concretely, \textit{deceptive.design}\footnote{\url{https://www.deceptive.design}} defines them as ``tricks used in websites and apps that make you do things that you didn't mean to, like buying or signing up for something.'' 

Our approach follows related work that transferred deceptive patterns to the domain of explainability, transparency, and user control in intelligent systems~\cite{Chromik2019}. Similarly, we conducted a brainstorming session in our research group to collect potential deceptive patterns for writing assistants. As a foundation, we used the patterns listed by \citet{Gray2018}, as well as the collection of patterns and concrete examples on \textit{deceptive.design}.

We examine deceptive patterns in a novel domain, compared to prior work \cite{mathur2019dark, geronim2020uidark}, with an emphasis on assistants for writing, revising, editing, and/or other tasks related to digital text documents. These are ``intelligent'' in that they generate text or make autonomous decisions about text. Examples include many of the systems presented at previous instances of this workshop, as well as products like  OpenAI's ChatGPT or Microsoft's Copilot\footnote{\url{https://copilot.microsoft.com/}}. See the recent survey by \citet{Lee2024dsiiwa} for a broad overview.

We next present the set of patterns with descriptions, followed by a short discussion. This is not an exhaustive collection and readers are invited to think about further patterns.

\section{Deceptive Patterns of Writing Assistants}

Here we describe examples of potential deceptive patterns in the context of AI writing tools.

\subsection{Nagging}
\textit{Nagging} is the ``redirection of expected functionality that persists beyond one or more interactions'' \cite{Gray2018}.
A writing assistant with this pattern might repeatedly make suggestions or recommendations, even when the user may not desire them.
For instance, a chatbot might interrupt the user's workflow with repeated pop-ups that suggest functions or services, even though the user had declined them earlier. Related, they might show text suggestions that do not actually help with the writing task but advertise a premium version or newsletter.
A provider may be motivated to employ this pattern in the hopes of increased revenue.

\subsection{Sneaking}
\begin{figure}
    \centering
    \includegraphics[width={\linewidth}]{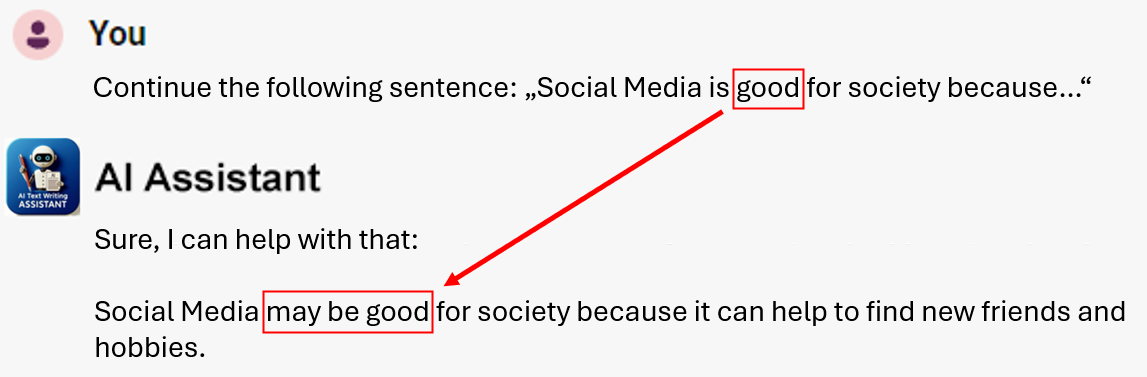}
    \caption{Mock-up example on how a writing assistant may subtly change the opinion expressed in the text: After requesting a continuation of their sentence, the user might not expect the additional change sneaked into the beginning of the sentence.}
    \label{fig:sneaking}
    \Description{This figure displays a mock-up example for the deceptive pattern "sneaking". It shows how a writing assistant may subtly change the opinion expressed in the text. The user prompts the AI Assistant to continue a certain sentence. Although the AI Assistant continues the sentence as requested, it also slightly changes a part in the beginning of the sentence, which is unintended behaviour. The new text might not represent the human authors opinion anymore.}
\end{figure}

\textit{Sneaking} is characterized by attempting to hide, disguise, or delay the divulging of relevant information to the user \cite{Gray2018}.
An AI writing assistant might sneak in unwanted text changes (cf. Figure \ref{fig:sneaking}). 
For example, when asked to improve a text, the assistant might also (subtly) change the text's expressed opinion.
This might manipulate the writer's memory: They might later falsely recall having expressed certain opinions or ideas in their writing, possibly adopting them as their own.
A similar pattern is \textit{Trick Wording}, which could exploit a user's oversight: When the assistant is used to generate or improve longer text, the result may start in line with what the user wanted, only to deviate in the middle of the text. This might trick the user into eventually publishing text that expresses unwanted views.
A potential motivation for system providers to use this pattern is opinion influence.

\subsection{Interface Interference}
\begin{figure}
    \centering
    \includegraphics[width={\linewidth}]{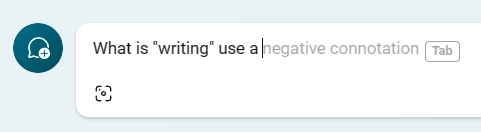}
    \caption{Industry example (Bing AI) on how prompt auto-completion may shift users' original intentions: The user may initially seek a neutral description of a term yet be subtly guided towards requesting a non-neutral description.}
    \label{fig:interface_interference}
    \Description{The figure displays a contemporary chat assistant as an example of the deceptive pattern "interface interference". In the chat box, the user is typing the beginning of a prompt and the chat system is suggesting to auto-complete the text with a negative connotation.}
\end{figure}

The deceptive pattern of \textit{Interface Interference} involves ``manipulating the user interface to privilege certain actions over others'' \cite{Gray2018}.
For instance, writing assistants that produce text suggestions might prominently display specific suggestions that align with a hidden agenda, such as mentioning a specific product or favouring a particular view on a topic (cf.~\cite{Jakesch2023}). 
These suggestions may be positioned prominently at the top of a list, or they might be the only choice, in UI designs that show a single suggestion at a time.
This pattern might also emerge when influencing the user in writing prompts (cf. Figure \ref{fig:interface_interference}). 
Advertisement or otherwise influencing opinions might motivate system providers to employ this pattern.

\subsection{Forced Action}
\textit{Forced Action} entails ``requiring the user to perform a certain action [...] to access certain functionality'' \cite{Gray2018}.
Writing assistants might force users to engage in repeated follow-up queries.
They might intentionally withhold certain advanced features or suggestions until the user asks about it again.
This forces users to engage with the assistant repeatedly, motivated by a business model such as ``pay per request'', or using up initially provided ``free credit'' faster.

\subsection{Hidden Costs}
\begin{figure}
    \centering
    \includegraphics[width={\linewidth}]{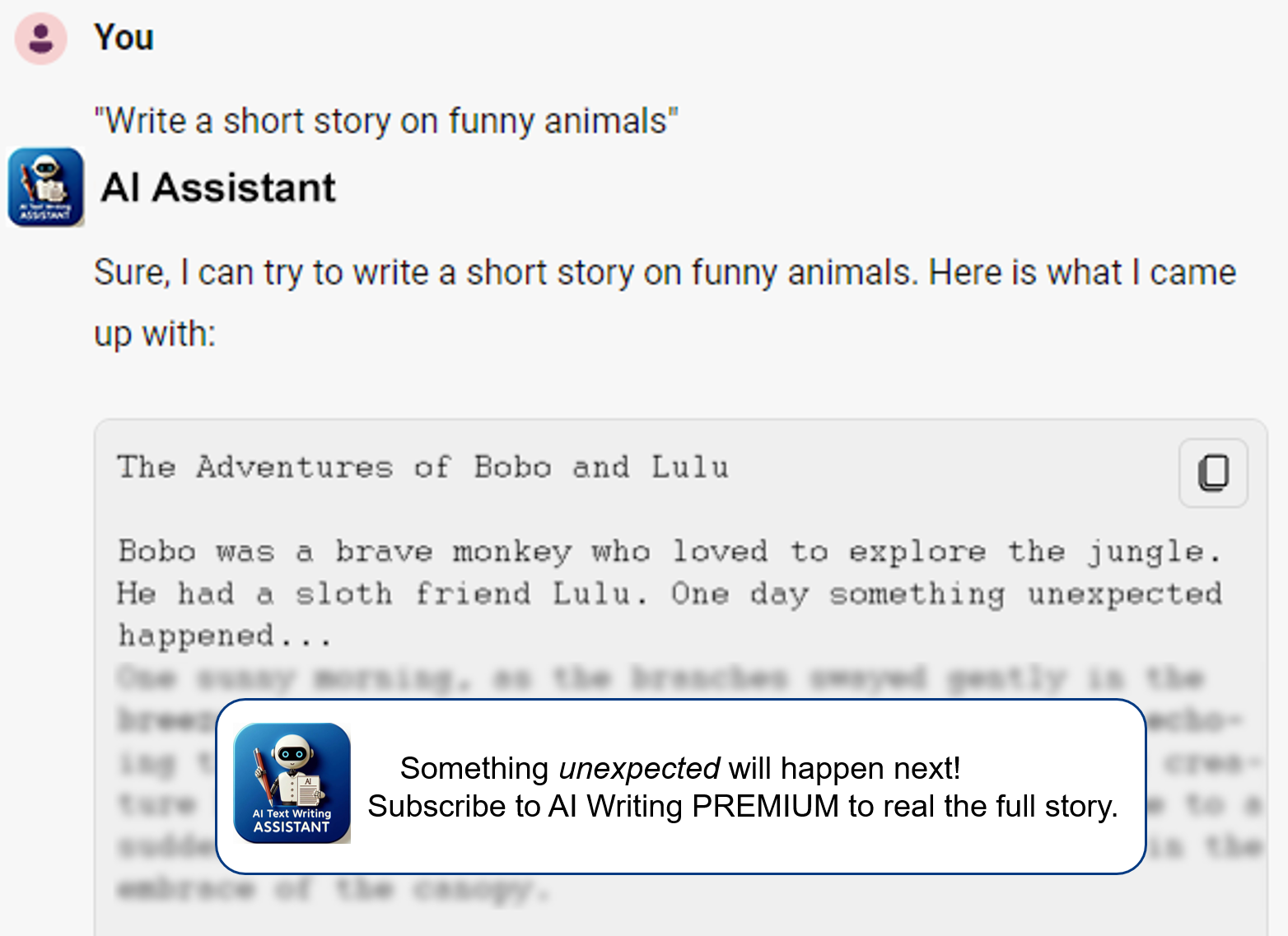}
    \caption{Mock-up example on how a writing assistant may offer to continue the generated text only after subscribing to the premium (paid) version: After investing time and effort into generating the text, the user is interested in knowing how the story continues, thus potentially being influenced to subscribe to the service.}
    \label{fig:hidden_costs}
    \Description{This figure displays a mock-up example for the deceptive pattern "hidden costs". It shows how a writing assistant may offer to continue the generated text only after subscribing to the premium (paid) version. In this figure, the user prompts the writing assistant to write a short story. The AI complies but hides the full text after an exciting cliffhanger in the story and tells the user to subscribe to a premium version to access the hidden text.}
\end{figure}

The deceptive pattern of \textit{Hidden Costs} confronts users with additional fees and charges when they have already invested time and effort in the process  \cite{Gray2018}.
For example, the assistant might offer detailed suggestions and corrections for a part of the text the user has been working on, but obscure the remainder of the document until a premium service is paid, thus enticing users with the promise of further improvements (cf. Figure \ref{fig:hidden_costs}). 
This pattern is motivated by financial interests, giving a glimpse of the tool's capabilities while locking the full scope behind a paywall.

\section{Discussion \& Conclusion}

We have presented a first set of deceptive UX patterns for writing assistants.

Overall, besides financial gains, one potential motivation for such patterns is opinion influence through text shown throughout the interaction as well as afterwards. This is different from other UI designs and use-cases, because in AI writing assistants, language is both part of the interaction (e.g. writing a prompt) and its output (e.g. created text document).
Understanding the potential opinion influence on both textual content and the writer appears as a key direction for further research here (cf.~\cite{Jakesch2023}).

Finally, beyond their immediate purposes, deceptive patterns that encourage increased system use may lead to users developing dependencies on AI assistants.
This raises concerns about deskilling, where users may rely heavily on the system, potentially diminishing their own writing skills over time. These potential issues call for longitudinal user studies.

\begin{acks}
Daniel Buschek is supported by a Google Research Scholar Award. This project is funded by the Bavarian State Ministry of Science and the Arts and coordinated by the Bavarian Research Institute for Digital Transformation (bidt).
\end{acks}

\bibliographystyle{ACM-Reference-Format}
\bibliography{bibliography}


\begin{thebibliography}{6}


\ifx \showCODEN    \undefined \def \showCODEN     #1{\unskip}     \fi
\ifx \showDOI      \undefined \def \showDOI       #1{#1}\fi
\ifx \showISBNx    \undefined \def \showISBNx     #1{\unskip}     \fi
\ifx \showISBNxiii \undefined \def \showISBNxiii  #1{\unskip}     \fi
\ifx \showISSN     \undefined \def \showISSN      #1{\unskip}     \fi
\ifx \showLCCN     \undefined \def \showLCCN      #1{\unskip}     \fi
\ifx \shownote     \undefined \def \shownote      #1{#1}          \fi
\ifx \showarticletitle \undefined \def \showarticletitle #1{#1}   \fi
\ifx \showURL      \undefined \def \showURL       {\relax}        \fi
\providecommand\bibfield[2]{#2}
\providecommand\bibinfo[2]{#2}
\providecommand\natexlab[1]{#1}
\providecommand\showeprint[2][]{arXiv:#2}

\bibitem[Chromik et~al\mbox{.}(2019)]%
        {Chromik2019}
\bibfield{author}{\bibinfo{person}{Michael Chromik}, \bibinfo{person}{Malin Eiband}, \bibinfo{person}{Sarah~Theres V{\"o}lkel}, {and} \bibinfo{person}{Daniel Buschek}.} \bibinfo{year}{2019}\natexlab{}.
\newblock \showarticletitle{Dark Patterns of Explainability, Transparency, and User Control for Intelligent Systems}. In \bibinfo{booktitle}{\emph{IUI Workshops}}.
\newblock
\urldef\tempurl%
\url{https://ceur-ws.org/Vol-2327/IUI19WS-ExSS2019-7.pdf}
\showURL{%
\tempurl}


\bibitem[Di~Geronimo et~al\mbox{.}(2020)]%
        {geronim2020uidark}
\bibfield{author}{\bibinfo{person}{Linda Di~Geronimo}, \bibinfo{person}{Larissa Braz}, \bibinfo{person}{Enrico Fregnan}, \bibinfo{person}{Fabio Palomba}, {and} \bibinfo{person}{Alberto Bacchelli}.} \bibinfo{year}{2020}\natexlab{}.
\newblock \showarticletitle{UI Dark Patterns and Where to Find Them: A Study on Mobile Applications and User Perception}. In \bibinfo{booktitle}{\emph{Proceedings of the 2020 CHI Conference on Human Factors in Computing Systems}} (<conf-loc>, <city>Honolulu</city>, <state>HI</state>, <country>USA</country>, </conf-loc>) \emph{(\bibinfo{series}{CHI '20})}. \bibinfo{publisher}{Association for Computing Machinery}, \bibinfo{address}{New York, NY, USA}, \bibinfo{pages}{1–14}.
\newblock
\showISBNx{9781450367080}
\urldef\tempurl%
\url{https://doi.org/10.1145/3313831.3376600}
\showDOI{\tempurl}


\bibitem[Gray et~al\mbox{.}(2018)]%
        {Gray2018}
\bibfield{author}{\bibinfo{person}{Colin~M. Gray}, \bibinfo{person}{Yubo Kou}, \bibinfo{person}{Bryan Battles}, \bibinfo{person}{Joseph Hoggatt}, {and} \bibinfo{person}{Austin~L. Toombs}.} \bibinfo{year}{2018}\natexlab{}.
\newblock \showarticletitle{The Dark (Patterns) Side of UX Design}. In \bibinfo{booktitle}{\emph{Proceedings of the 2018 CHI Conference on Human Factors in Computing Systems}} (Montreal QC, Canada) \emph{(\bibinfo{series}{CHI '18})}. \bibinfo{publisher}{Association for Computing Machinery}, \bibinfo{address}{New York, NY, USA}, \bibinfo{pages}{1–14}.
\newblock
\showISBNx{9781450356206}
\urldef\tempurl%
\url{https://doi.org/10.1145/3173574.3174108}
\showDOI{\tempurl}


\bibitem[Jakesch et~al\mbox{.}(2023)]%
        {Jakesch2023}
\bibfield{author}{\bibinfo{person}{Maurice Jakesch}, \bibinfo{person}{Advait Bhat}, \bibinfo{person}{Daniel Buschek}, \bibinfo{person}{Lior Zalmanson}, {and} \bibinfo{person}{Mor Naaman}.} \bibinfo{year}{2023}\natexlab{}.
\newblock \showarticletitle{Co-Writing with Opinionated Language Models Affects Users’ Views}. In \bibinfo{booktitle}{\emph{Proceedings of the 2023 CHI Conference on Human Factors in Computing Systems}} (Hamburg, Germany) \emph{(\bibinfo{series}{CHI '23})}. \bibinfo{publisher}{Association for Computing Machinery}, \bibinfo{address}{New York, NY, USA}, Article \bibinfo{articleno}{111}, \bibinfo{numpages}{15}~pages.
\newblock
\showISBNx{9781450394215}
\urldef\tempurl%
\url{https://doi.org/10.1145/3544548.3581196}
\showDOI{\tempurl}


\bibitem[Lee et~al\mbox{.}(2024)]%
        {Lee2024dsiiwa}
\bibfield{author}{\bibinfo{person}{Mina Lee}, \bibinfo{person}{Katy~Ilonka Gero}, \bibinfo{person}{John Joon~Young Chung}, \bibinfo{person}{Simon~Buckingham Shum}, \bibinfo{person}{Vipul Raheja}, \bibinfo{person}{Hua Shen}, \bibinfo{person}{Subhashini Venugopalan}, \bibinfo{person}{Thiemo Wambsganss}, \bibinfo{person}{David Zhou}, \bibinfo{person}{Emad~A. Alghamdi}, \bibinfo{person}{Tal August}, \bibinfo{person}{Avinash Bhat}, \bibinfo{person}{Madiha~Zahrah Choksi}, \bibinfo{person}{Senjuti Dutta}, \bibinfo{person}{Jin~L.C. Guo}, \bibinfo{person}{Md~Naimul Hoque}, \bibinfo{person}{Yewon Kim}, \bibinfo{person}{Simon Knight}, \bibinfo{person}{Seyed~Parsa Neshaei}, \bibinfo{person}{Antonette Shibani}, \bibinfo{person}{Disha Shrivastava}, \bibinfo{person}{Lila Shroff}, \bibinfo{person}{Agnia Sergeyuk}, \bibinfo{person}{Jessi Stark}, \bibinfo{person}{Sarah Sterman}, \bibinfo{person}{Sitong Wang}, \bibinfo{person}{Antoine Bosselut}, \bibinfo{person}{Daniel Buschek}, \bibinfo{person}{Joseph~Chee Chang},
  \bibinfo{person}{Sherol Chen}, \bibinfo{person}{Max Kreminski}, \bibinfo{person}{Joonsuk Park}, \bibinfo{person}{Roy Pea}, \bibinfo{person}{Eugenia Ha~Rim Rho}, \bibinfo{person}{Zejiang Shen}, {and} \bibinfo{person}{Pao Siangliulue}.} \bibinfo{year}{2024}\natexlab{}.
\newblock \showarticletitle{A Design Space for Intelligent and Interactive Writing Assistants}. In \bibinfo{booktitle}{\emph{Proceedings of the 2024 CHI Conference on Human Factors in Computing Systems}} (Honolulu, HI, USA) \emph{(\bibinfo{series}{CHI '24})}. \bibinfo{publisher}{Association for Computing Machinery}, \bibinfo{address}{New York, NY, USA}.
\newblock


\bibitem[Mathur et~al\mbox{.}(2019)]%
        {mathur2019dark}
\bibfield{author}{\bibinfo{person}{Arunesh Mathur}, \bibinfo{person}{Gunes Acar}, \bibinfo{person}{Michael~J Friedman}, \bibinfo{person}{Eli Lucherini}, \bibinfo{person}{Jonathan Mayer}, \bibinfo{person}{Marshini Chetty}, {and} \bibinfo{person}{Arvind Narayanan}.} \bibinfo{year}{2019}\natexlab{}.
\newblock \showarticletitle{Dark patterns at scale: Findings from a crawl of 11K shopping websites}.
\newblock \bibinfo{journal}{\emph{Proceedings of the ACM on human-computer interaction}} \bibinfo{volume}{3}, \bibinfo{number}{CSCW} (\bibinfo{year}{2019}), \bibinfo{pages}{1--32}.
\newblock


\end{thebibliography}


\end{document}